\documentclass[preprint,12pt,authoryear]{elsarticle}

\usepackage{amssymb}

\usepackage{orcidlink}
\usepackage{mathtools}
\usepackage[utf8]{inputenc}
\usepackage[T1]{fontenc}

\definecolor{amethyst}{rgb}{0.6, 0.4, 0.8}
\definecolor{green}{rgb}{0.55, 0.71, 0.0}
\definecolor{apricot}{rgb}{0.98, 0.81, 0.69}
\definecolor{auburn}{rgb}{0.43, 0.21,0.1}
\definecolor{babyblueeyes}{rgb}{0.63, 0.79, 0.95}
\definecolor{bittersweet}{rgb}{1.0, 0.44, 0.37}
\definecolor{blue(munsell)}{rgb}{0.0, 0.5, 0.69}
\definecolor{oceanboatblue}{rgb}{0.0, 0.47, 0.75}
\definecolor{brightmaroon}{rgb}{0.76, 0.13, 0.28}
\definecolor{deepcarminepink}{rgb}{0.94, 0.19, 0.22}

\journal{Nuclear Instruments and Methods in Physics Research - section A (NIM-A)}

\begin{document}

\begin{frontmatter}



\title{Performance Evaluation of Three Silicon Photomultiplier Detector Modules within the MAGIC Telescopes PMT-based camera}


\author[a]{A. Hahn\corref{cor1}\orcidlink{0000-0003-0827-5642}}
\ead{ahahn@mpp.mpg.de}
\cortext[cor1]{Corresponding author}

\author[a]{R. Mirzoyan\orcidlink{0000-0003-0163-7233}}
\ead{razmik.mirzoyan@mpp.mpg.de}

\author[a]{A. Dettlaff}
\ead{todettl@mpp.mpg.de}

\author[a]{D. J. Fink}
\ead{fink@mpp.mpg.de}

\author[a,b]{D. Mazin\orcidlink{0000-0002-2010-4005}}
\ead{mazin@icrr.u-tokyo.ac.jp}

\author[a,b]{M. Teshima}
\ead{mteshima@icrr.u-tokyo.ac.jp}

\address[a]{Max Planck Institute for Physics,
            { Föhringer Ring 6}, 
            Munich,
            D-80805, 
            Bavaria,
            Germany}
            
\address[b]{Institute for Cosmic Ray Research, The University of Tokyo,
            { 5-1-5 Kashiwa-no-Ha}, 
            Kashiwa City,
            277-8582, 
            Chiba,
            Japan}

\begin{abstract}

MAGIC is a system of two imaging atmospheric Cherenkov telescopes (IACTs) located on the Canary island of La Palma. Each telescope's imaging camera consists of 1039 photomultiplier tubes (PMTs). We developed three detector modules based on silicon photomultipliers (SiPMs) of seven pixels each that are mechanically and electronically compatible with those used in the MAGIC camera. These prototype modules are installed next to the PMTs in the imaging camera and are operated in parallel. To achieve a similar active area per pixel we used seven to nine SiPMs for producing a composite pixel. The SiPM signals within one such pixel are actively summed up for retaining the fast signal pulse shapes. Two different PCB designs are tested for thermal performance. We present our simulations of Cherenkov and light of the night sky (LoNS) responses. Based on those we calculate the signal-to-noise ratio (SNR) for this imaging application. We compare our expectations with the measurements of one of the SiPM-based detector modules.

\end{abstract}









\end{frontmatter}


\section{Introduction}
\label{sec:introduction}
\hspace{15mm} Imaging Atmospheric Cherenkov Telescopes (IACTs) observe the Cherenkov light emitted by a cascade of secondary particles, the so called extensive air shower, created by the interaction of a primary cosmic and gamma rays with the atoms of the atmosphere. These Cherenkov light flashes are faint and last only few nanoseconds \citep{jelley_cerenkov_1955}. All cameras of large-size IACTs are currently based on PMTs as light detectors \citep{mirzoyan_technological_2022, aharonian_observations_2006, mazin_large_2016, rajotte_upgrade_2014,borwankar_estimation_2020}. Smaller IACTs like the First G-APD Cherenkov Telescope (FACT) or the prototype Schwarzschild-Couder telescopes of the Small-Sized and Medium-Sized Telescope (SST, MST) type of the Cherenkov Telescope Array (CTA) use Silicon Photomultipliers (SiPMs) as light detectors \citep{krahenbuhl_g-apds_2012,mirzoyan_sipm_2013,montaruli_small_2015,benbow_status_2017,catalano_astri_2018}. The increasing popularity of SiPMs might suggest that also large-size IACTs could use SiPMs, however, there is no conclusive study that directly compares PMT and SiPM performances for IACTs. Such comparison is not straightforward because of the presence of the light of the night sky (LoNS) as a strong background for the IACT technique. 

The MAGIC (Major Atmospheric Gamma Imaging Cherenkov) telescopes are located in the Roque de los Muchachos observatory on the Canary island of La Palma. The imaging cameras of MAGIC offer the possibility to install up to six additional prototype detector modules of seven pixels each, next to the PMT-based ones. We developed three SiPM prototype detector modules in our institute. These modules are adapted to the mechanical, electronic and signal requirements for installation into the MAGIC-1 camera. We developed calibration procedures, checked the modules stability and compared the performances using the light flashes from the calibration laser as well as from air showers during telescope operations.

\section{Module design}
\label{sec:module_design}
The light-sensitive imaging surface of the MAGIC camera is circular in shape and consists of 1039 1-inch size PMT pixels of type Hamamatsu R10408. These are grouped in modules that hold up to seven pixels \citep{aleksic_major_2016}. The mechanical structure of the camera is hexagonal \citep{nakajima_new_2013}. The used circular shape of the PMT arrangement in the camera leaves the six vertices of the hexagon free. In those six locations, prototype detector modules can be installed without interfering with the scientific operation of the telescopes. Each pixel in the imaging camera is given a hexagonal-shape light guide on top that minimizes the dead area of the camera and rejects the stray light.\\
In order to emulate a 1-inch PMT pixel surface area we used matrices of seven or nine SiPMs of $6\times6\,\mathrm{mm}^2$ size, topped by a dedicated light guide described in \citet{hahn_development_2017}. The individual SiPM signals have to be summed to form a combined pixel output signal. Because the light guide is non-imaging and the optical point-spread-function of the telescope comparable to pixel size, the readout of the individual SiPMs does not provide any additional knowledge and only increases complexity and cost.\\
We use a discrete transistor common base circuits with low input impedance and high output impedance to sum the currents of the seven or nine SiPMs for producing one composite pixel. This allows one to avoid adding up the device capacitances, which otherwise would worsen the fast signal response times necessary for the application in IACTs. The circuit design and simulation is described in \citet{fink_sipm_2016}. Each SiPM pixel also has a temperature sensor on the front side, right next to the SiPMs.\\
We installed the first SiPM-based detector module in May 2015 in the MAGIC-1 camera. In the first generation pixels we used seven Excelitas C30742-66-050-X SiPMs. Later on we used our experience gained with this first generation to design an improved second generation SiPM-based detector modules, which utilize nine SiPMs per pixel. One of the second generation modules uses Hamamatsu S13360-6075VS SiPMs, and the other one uses SensL MicroFJ-60035-TSV. These Hamamatsu and SensL modules were installed in the MAGIC-1 camera in 2017. While for the first generation SiPM pixels we used regular printed circuit boards (PCBs), for the second generation we used PCBs with a thick Aluminum layer sandwiched in the center for improving the heat conductivity from the SiPMs to the Aluminum structure of the detector module.\\
All SiPM modules are integrated into the standard MAGIC slow control software and are operated in parallel with the PMT camera. The SiPM modules are read out by the standard MAGIC data acquisition system, described in \citet{sitarek_analysis_2013}. The MAGIC trigger system does not include all PMT pixels but only an inner region of 547 pixels \citep{aleksic_major_2016}. This means that our SiPM modules operate in a parasitic trigger mode on events that trigger the inner region of the camera.

\section{Temperature stability}
\label{sec:temperature}
\begin{figure}
    \centering
    \includegraphics[width=\columnwidth]{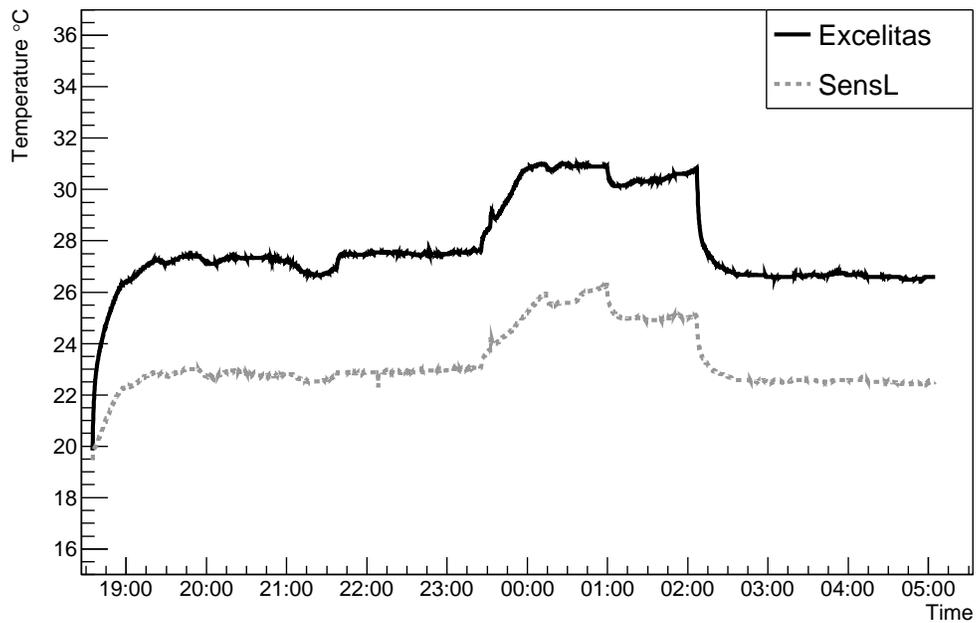}
    \caption{Measured temperatures of one first (Excelitas) and one second generation (SensL) SiPM pixel. After an initial warm-up phase, the data taking is started with a stable pixel temperature. Around midnight the bright moon rises, creating a lot of background light. This causes the pixels to heat up. Shortly past 2\,a.m., the camera is closed and the temperature drops to a level almost corresponding to that of the dark night.}
    \label{fig:temperature}
\end{figure}

\begin{figure}
    \centering
    \includegraphics[width=\columnwidth]{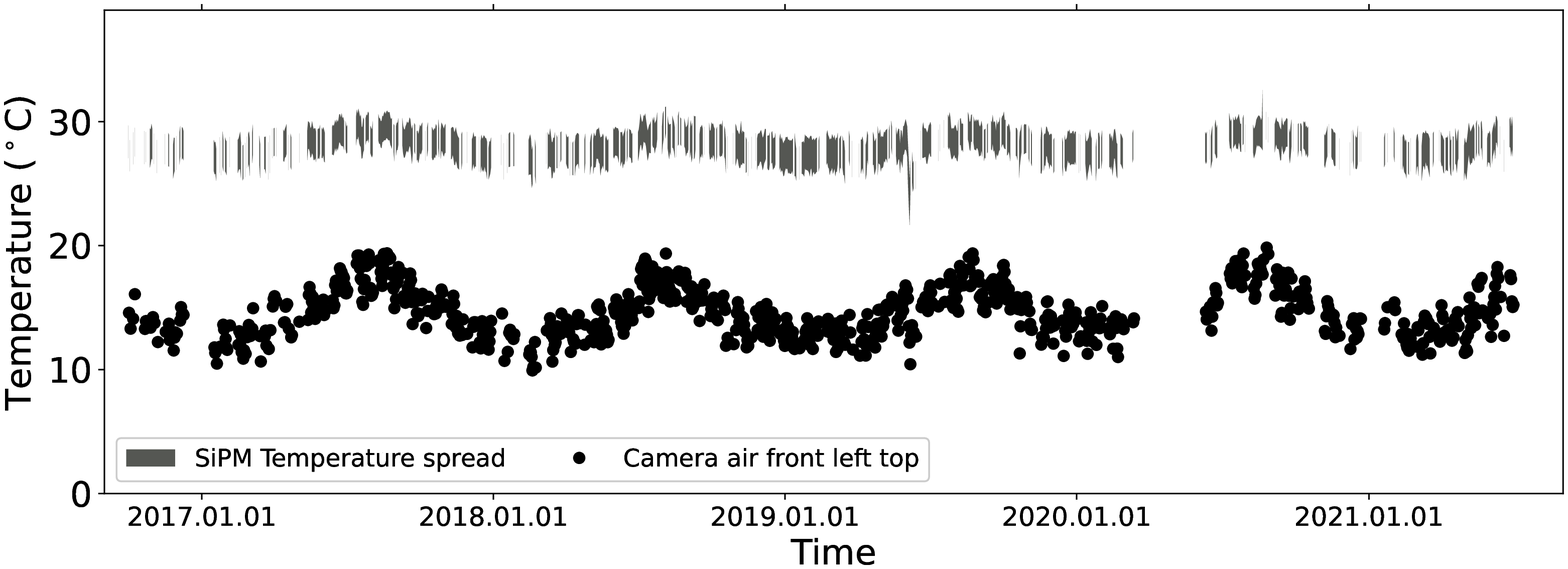}
    \caption{Measured SiPM pixel temperature. For every night the 2-sigma range of pixel temperatures is shown. This excludes high current transients from passing stars which have to be excluded from the gamma-ray analysis as well. Special test runs with different SiPM voltage configurations were excluded. The camera temperature measured at the front top in the air close to the SiPM module is shown as well.}
    \label{fig:long_term_temperature}
\end{figure}

The temperature of the MAGIC cameras are actively controlled via a closed-loop circulation of a cooling liquid \citep{nakajima_new_2013}. Both generations of SiPM pixels show a stable operational temperature after an initial warm-up phase during the twilight, see figure \ref{fig:temperature}. One can see that the pixels with Aluminum core PCB reach stable operation at $5^\circ$C lower temperature than first generation pixels. During the moon rise around midnight in this example, the increasing background light induces a higher current, which heats up the SiPM pixels. It can be seen that with the rising moon the pixel temperatures increase and then stabilize at the new level. When the camera is closed, the temperature drops to a value $\sim1^\circ$C below the pixel temperature during dark night observations.\\
The SiPM pixel temperature during more than 4 years of operation was stable within $\pm 4^\circ$C, see figure \ref{fig:long_term_temperature}. One can see that the seasonal pixel temperature variations are significantly less than the temperature variations of the air inside the camera.

\section{Calibration}
\label{sec:calibration}
We use two methods for the absolute charge-to-photoelectron calibration of the SiPMs signals. The first one is based on the charge distribution of presumably Poisson-distributed dark counts. With this, one can directly estimate the calibration constant and cross-talk probability from the separated peaks of the charge histogram. The charge extraction and event selection algorithms are described in \citet{hahn_development_2018}.\\
The second method for calibration is the so called F-factor (also known as excess noise factor) method. This direct and reliable way is routinely used for the MAGIC camera for calibrating the PMT-pixel gains. It also has the advantage of not requiring single photoelectron peaks resolved in amplitude. The precision of the formalism depends on the skills of the experimenter to extend the extraction of the spectrum towards the lowest charge genuine single photoelectron events, see \citet{mirzoyan_calibration_1997}. For PMTs, the F-factor can be determined in laboratory measurements. For SiPMs, the main contribution to the excess noise factor is the crosstalk \citep{vinogradov_analytical_2012}. So the cross talk probability of the SiPM pixel has to be determined first to be able to apply the F-factor calibration method.



For the F-factor calibration method in MAGIC we use light flashes of the third harmonic of a Nd:YAG laser (355\,nm). This illuminates the camera homogeneously from 17m distance. These light flashes can not only be used for calibration but also for the first performance check. Considering the corresponding wavelength-dependent photon detection efficiency (PDE) and the percentage of the light-sensitive area of a composite pixel, one can calculate the expected number of photoelectrons for the different SiPM pixels. For the Excelitas based first generation pixel design we calculate a relative efficiency of 50\% compared to the MAGIC PMTs and $\sim$100\% for both of the second generation pixel types. These estimates were confirmed by our measurements which can serve as a validation of our SiPM calibration procedures.

\section{Cherenkov light and LoNS}
\label{sec:cherenkov_lons}
\begin{figure}
    \centering
    \includegraphics[width=\columnwidth]{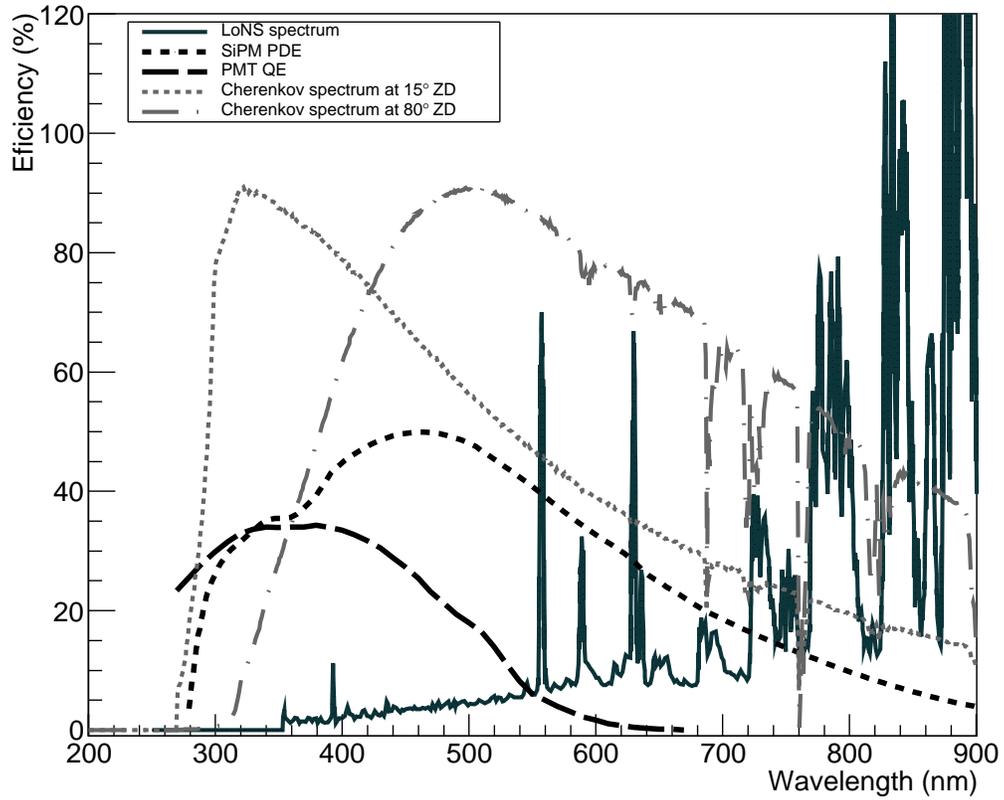}
    \caption{The LoNS spectrum, taken from \citet{benn_palma_1998}, is shown as solid line. The Cherenkov spectrum for an extensive air shower at an incident angle of $15^\circ$ ZD is shown as fine dotted line and that for an incident angle of $80^\circ$ ZD as dash-dotted line. The LoNS and Cherenkov spectra were scaled for better visibility. The MAGIC PMT QE curve is shown as long-dashed line and the Hamamatsu SiPMs' PDE is shown as black dotted line.}
    \label{fig:spectra}
\end{figure}

The amount of Cherenkov light from an extensive air shower that reaches the telescopes depends on the observation zenith angle. It is due to light extinction in the atmosphere along the different path lengths. At a low zenith distance (ZD), the Cherenkov spectrum peaks at $\sim320$\,nm. With increasing ZD the peak gets shifted towards the green due to the extinction of the near UV and blue photons. For our study we simulated the air shower Cherenkov spectra with CORSIKA \citep{heck_corsika_1998} and the atmosphere, the mirror and the imaging camera geometry with the MAGIC Analysis and Reconstruction Software (MARS) \citep{zanin_mars_2013,aleksic_major_2016-2}. We expanded the atmospheric simulation in MARS by increasing the range of simulated wavelength and inclusion of the most dominant molecular absorption lines from the high-resolution transmission molecular absorption database (HITRAN)\footnote{\url{https://hitran.org/}} \citep{gordon_hitran2020_2022}. Two example Cherenkov spectra for 5\,TeV primary gamma-ray air showers incident at $\mathrm{ZD}=15^\circ$ and $\mathrm{ZD}=80^\circ$ are shown in figure \ref{fig:spectra}. We simulated 5 and 50\,TeV gamma-ray events which are far above the set trigger threshold of the MAGIC telescopes \citep{aleksic_major_2016}.\\
The measured spectrum of LoNS at La Palma starts at around 350\,nm and increases towards longer wavelengths. The measurements, reported in \citet{benn_palma_1998}, were performed at low ZD; one example is shown in figure \ref{fig:spectra}. One of the main contributions to the LoNS is the airglow which will increase with increasing airmass due to the longer projected airglow layer thickness \citep{van_rhijn_brightness_1919, van_rhijn_brightness_1921}. At very large ZDs, parts of the light from airglow are again reduced because the van Rhijn effect can no longer compensate for the atmospheric extinction of some wavelength \citep{hong_transfer_1998, noll_atmospheric_2012}. 
Since atmospheric extinction is wavelength dependent, the change in airglow with ZD is also wavelength dependent. We therefore take the low ZD measurement on La Palma by \citet{benn_palma_1998} and extrapolate each wavelength using the relative changes with ZD simulated with the \textit{SKYCALC Sky Model Calculator}\footnote{\url{https://www.eso.org/observing/etc/bin/gen/form?INS.MODE=swspectr+INS.NAME=SKYCALC}} using the Cerro Paranal Sky Model \citep{noll_atmospheric_2012,jones_advanced_2013}.\\
To measure the detection efficiency, for one night we installed the SiPM-based module equipped with Hamamatsu SiPMs at the center of the MAGIC-1 camera. In the following we will  focus on the performance of this SiPM-based module.\\
The quantum efficiency (QE) of the MAGIC PMTs closely matches the Cherenkov light emission spectrum from low ZD air showers. It is not sensitive at long wavelengths where the LoNS spectrum is steeply increasing, see figure \ref{fig:spectra}. The PDE of Hamamatsu SiPMs reaches its maximum at longer wavelengths than that of PMT. Moreover, it shows a long tail. This means that, in general, SiPMs will detect less of the near UV and blue but more of the green part of the Cherenkov light, which suggests that SiPMs could have an advantage for observations at large and very large ZDs. However, the long-wavelength sensitivity tail also leads to a much increased LoNS background rate. 
By multiplying the wavelength-dependent detection efficiency of the SiPMs and the QE and the photoelectron collection efficiency of the PMTs with the LoNS and Cherenkov spectra one can calculate and compare the expected performances. For observations at low ZD we calculated that the Hamamatsu SiPMs will detect 1.8 times more Cherenkov photons but also 4.9 times more LoNS photons compared to the MAGIC PMTs.\\
Here we would like to remind that the PMTs used in MAGIC are more than a decade old. Today, better PMTs with $\sim30\%$ higher QE and other strongly improved parameters (see for example \citet{toyama_evaluation_2015,mirzoyan_evaluation_2017}) but also newer SiPMs are available.\\
To test the above estimates, we analyzed the gathered 40 minutes of dark data by selecting only very large shower images that illuminate the entire center of the camera so that SiPM and neighboring PMT pixels are illuminated simultaneously. We calibrated the Hamamatsu SiPMs using the aforementioned dark count single photoelectron histogram method and cross-checked the result with the F-factor method. The results essentially confirmed our calculations with measured values of 4.3 times more LoNS and 1.9 times more Cherenkov photons. The $\sim12\%$ difference in LoNS intensity still could be considered in agreement because the LoNS not only depends on the simulated parameters but also on the rotating star field in the imaging camera plane, variable atmospheric conditions \citep{fruck_characterizing_2022} and variations in the LoNS \citep{roach_characteristic_1958,patat_dancing_2008}.

\section{Signal-to-noise ratio}
\label{sec:snr}

\begin{figure}
    \centering
    \includegraphics[width=\columnwidth]{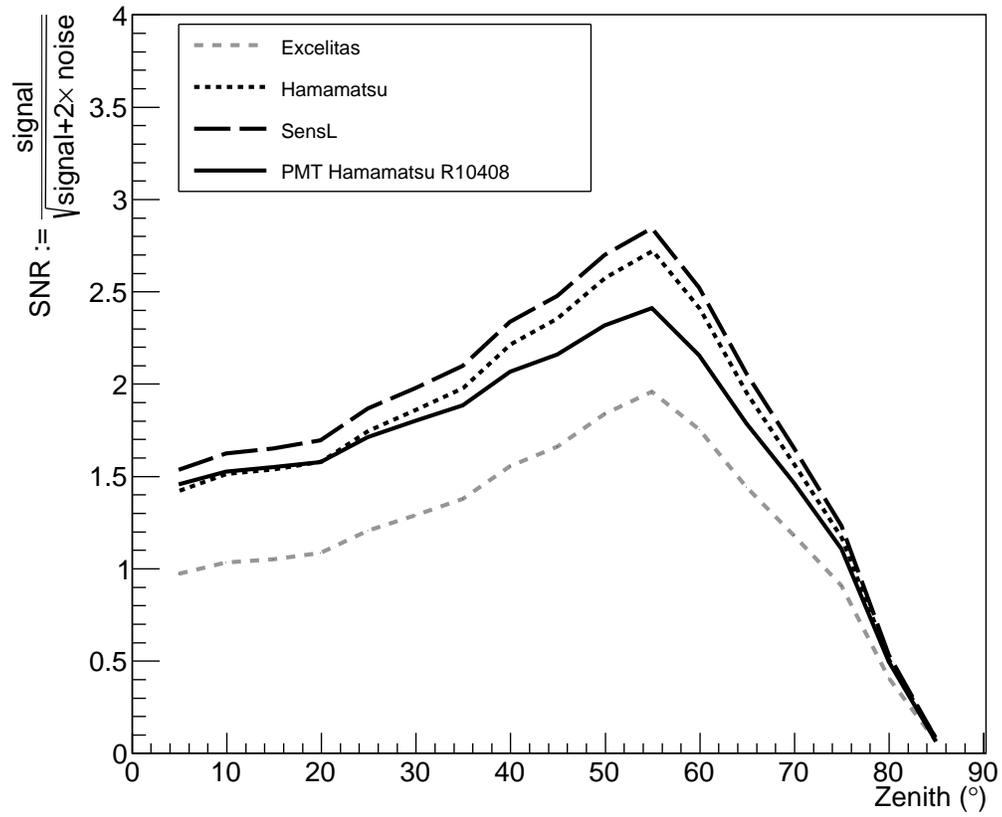}
    \caption{Simulated SNR of the MAGIC PMTs and the used Excelitas, Hamamatsu and SensL SiPMs of our prototype detector modules inside the MAGIC-1 camera versus observation zenith angle.}
    \label{fig:snr}
\end{figure}

In imaging applications the Signal-to-noise ratio (SNR) is defined as the ratio of the mean signal $\mu_{\mathrm{signal}}$ to the standard deviation of the noise $\sigma_\mathrm{noise}$. Since both the signal $S$ and the noise $N$ are Poisson-distributed it follows that\\
\begin{equation}
    \mathit{SNR} \coloneqq \frac{\mu_{\mathrm{signal}}}{\sigma_\mathrm{noise}} = \frac{\langle S \rangle }{\sqrt{\langle S \rangle  + \langle N \rangle}}.
\label{eq:snr}
\end{equation}
A measurement of $S$ always contains the real signal $S_\mathrm{true}$ and a noise contribution that can not be disentangled:
\begin{equation}
\langle S \rangle = \langle S_\mathrm{true} + N \rangle = \langle S_\mathrm{true} \rangle + \langle N \rangle.
\label{eq:signal}
\end{equation}
The \textit{SNR} of a background subtracted measurement is therefore
\begin{equation}
    \mathit{SNR} = \frac{\langle S_\mathrm{true}\rangle}{\sqrt{\langle S_\mathrm{true} \rangle + 2\times\langle N \rangle}}.
\label{eq:snr_trigger}
\end{equation}
The Monte-Carlo simulated Cherenkov air shower correspond to $S_\mathrm{true}$ and the calculated LoNS to $N$ so that the \textit{SNR} can be calculated directly. The \textit{SNR} versus ZD is shown in figure \ref{fig:snr} for the simulated fixed energy gamma-ray events. One can see that although these events are far above the trigger threshold there is no clear benefit of modern SiPMs over the (old) MAGIC PMTs. At larger ZDs the Hamamatsu and SensL SiPM somewhat outperform the MAGIC PMTs due to their higher PDE at longer wavelength (see figure \ref{fig:spectra}). At very large ZDs, due to large involved distances, the shower images in the telescope camera become very small while the LoNS increases, both reducing the SNR difference.

\section{Summary}
\label{sec:summary}
We developed three SiPM-based detector modules to measure the signal-to-noise ratio of large-size composite pixels in real IACT operations. These include seven SiPMs in the first generation and nine SiPMs in the second generation. The 7--9 SiPMs in one pixel are summed up by using transimpedance transistor circuits, one per SiPM chip, to preserve the fast signal shape. We utilized two different calibration procedures. Our three SiPM-based modules have been operated since 2015/2017 and show stable multi-year behavior of calibration and seasonal temperature changes. The simulated and measured LoNS and Cherenkov light responses agree within uncertainties. We calculated the SNR for SiPM and PMT-based pixel for zenith angles up to $85^\circ$. For low ZD observations the SiPM pixels perform marginally better than the MAGIC PMTs. For observations at large zenith angles the SiPMs somewhat outperform the MAGIC PMTs due to their increased sensitivity to longer wavelength photons. For very large zenith observations, MAGIC PMTs and SiPMs perform again very similar due to smaller showers and increasing LoNS.

In near future we plan to replace the extrapolated LoNS spectrum used in this study by new LoNS measurements performed at the MAGIC site. This will allow us to overcome the limitations of the current van Rhijn modeling at very large zenith angles described in \citet{g_t_best_new_1965,noll_atmospheric_2012}. An obvious proposal for SiPM-based imaging cameras is to use UV/blue-pass filters to reduce the effect of the LoNS (see e.g.~\citet{montaruli_small_2015,catalano_astri_2018}). We plan to study this in simulation and in measurement. By measuring the trigger rates for different thresholds we plan to measure the energy threshold of an IACT camera fully equipped with the SiPM-based pixels.

\section{Acknowledgments}
\label{sec:acknowledgments}

We gratefully acknowledge the support of the MAGIC collaboration, in particular by the local La Palma team, the various shift crews, the hardware and software experts and the safety and operations committee. 
The projects seed money was provided by the Otto-Hahn award of the Max Planck Society to Daniel Mazin.  We would like to thank the Instituto de Astrof\'{\i}sica de Canarias for the excellent working conditions at the Observatorio del Roque de los Muchachos in La Palma.
This work was financially supported by the German MPG and the Japanese ICRR, the University of Tokyo.



 \bibliographystyle{elsarticle-num-names} 

\bibliography{bibliography.bib}





\end{document}